\documentclass{PoS}
\usepackage{amsmath,amssymb,graphics,epsfig}
\usepackage{mcite}
\newcommand{\hatA}[0]{\hat{A}}

\newcommand{\tr}[0]{\text{tr}}

\newcommand{\ket}[1]{\left|#1\right\rangle}
\newcommand{\bra}[1]{\left\langle#1\right|}

\newcommand{\qed}{\nobreak \ifvmode \relax \else
      \ifdim\lastskip<1.5em \hskip-\lastskip
      \hskip1.5em plus0em minus0.5em \fi \nobreak
      \vrule height0.75em width0.5em depth0.25em\fi}

\usepackage{grffile}

\title{String tension from gauge invariant Magnetic Monopoles}

\ShortTitle{String tension from gauge invariant Magnetic Monopoles}

\author{\speaker{Nigel Cundy}$^a$, Weonjong Lee$^a$, Jaehoon Leem$^a$ and Y. M. Cho$^b$%
\\
    \llap{$^a$}    Lattice Gauge Theory Research Center, FPRD, and CTP, Department of Physics \&
    Astronomy,\\ Seoul National University, Seoul, 151-747, South Korea\\
\llap{$^b$} 
Administration Building 310-4, Konkuk University,
Seoul 143-701, Korea

  E-mail: \email{ndcundy@phya.snu.ac.kr}
}

\abstract{We investigate the relationship between colour confinement and the monopoles derived from the Cho-Duan-Ge decomposition. These monopoles, unlike Dirac and 't Hooft monopoles, do not require a singular gauge field and are defined for any choice of gauge (and are not just restricted to, for example, the maximum Abelian gauge). The Abelian decomposition is defined in terms of a colour field $n$; the principle novelty of our study is that we have used a unique definition of this field in terms of the eigenvectors of the Wilson Loop. This allows us to investigate the relationship between the gauge invariant monopoles and confinement both analytically and numerically, as well as retaining the maximal possible symmetry within the colour field so that it is able to see all the monopoles in an SU($N_C$) calculation.

We describe how the Abelian decomposition is related to the Wilson Loop, so that the string tension may be calculated from the field strength related to the decomposed (or restricted) Abelian field. We demonstrate that for an area law scaling of the Wilson Loop there must be discontinuities in the restricted field, and discuss the structures in the colour field which may cause these discontinuities, which turn out to be magnetic monopoles. If these monopoles are present, they will lead to an area law scaling of the Wilson Loop and thus be at least partially responsible for confinement.

We search for these monopoles in quenched lattice QCD. We show that the string tension is dominated by peaks in the restricted field strength, at least some of which are located close to structures in the colour field consistent with with theoretical expectations for the monopoles. We show that the string tension extracted from the monopole contribution to the restricted field is close to that of the entire original field; again suggesting that confinement can at least partially be explained in terms of these monopoles.
}

\FullConference{The 30th International Symposium on Lattice Field Theory\\
                 June 24 -- 29,  2012\\
                 Cairns, Australia}

\begin{document}

\section{Introduction}

The mechanism behind quark confinement is an enduring problem in QCD. Although several models have been proposed -- for example, center vortices~\cite{Vortices}, and a dual Meissner effect due to magnetic monopoles~\cite{Monopoles1,*Monopoles2,*Mandelstam:1976,*thooft:1976} -- none have yet been demonstrated convincingly. Here we study the possibility of confinement due to monopoles constructed from the Cho-Duan-Ge (CDG) decomposition (sometimes referred to as the Cho-Faddeev-Niemi decomposition)~\cite{Cho:1980,*Cho:1981,*F-N:98,*Shabanov:1999,*Duan:1979}. Unlike Dirac monopoles and 't Hooft (Maximum Abelian Gauge) monopoles, the CDG decomposition respects the gauge symmetry and does not require a singular gauge field. The decomposition is constructed from a colour field, $n$, which is built from a matrix $\theta$. Recent work~\cite{Kondo:2008su,*Shibata:2009af,*Kondo:2010pt,*Kondo:2005eq} has demonstrated that the monopole contribution dominates the confining string, using monopoles constructed from one 
particular choice of $\theta \in
SU(N_C)/U(N_C-1)$; however in this case only one
of the possible $N_C-1$ types of monopole is visible. Here we consider a different choice of $\theta \in SU(N_C)/(U(1))^{N_C-1}$, and investigate whether this the monopoles apparent in this construction may also lead to confinement. Our full results and methods will be published later~\cite{CundyForthcoming}. In section \ref{sec:2} we discuss the Abelian decomposition and its relation to the Wilson Loop and thus quark potential; in section \ref{sec:3} we discuss how monopoles may arise in this construction and lead to confinement; we present numerical evidence in section \ref{sec:4} and conclude in section \ref{sec:5}.

\section{Abelian decomposition and Stokes' theorem}\label{sec:2}
The confining potential in an SU($N_C$) gauge theory can be measured using the Wilson Loop,
\begin{align}
W_L[C_s] = & \frac{1}{N_C} \tr \left(W[C_s]\right) & W[C_s] = P[e^{-ig\oint_{C_s} dx_\mu A_\mu(x)}]
\end{align}
for a closed curve $C_s$ of length $L$ which starts and finishes at a position $s$, where $P$ represents path ordering and the gauge field, $A_\mu$, can be written in terms of the Gell-Mann matrices, $\lambda^a$, as $A_\mu^a \lambda^a$. The expectation value of the Wilson Loop scales as $\langle W_L[C_s]\rangle \sim e^{-\mathfrak{s} \mathcal{A}}$, where $\mathcal{A}$ is the area of the surface enclosed by the curve $C_s$ and $\mathfrak{s}$ is the string tension. 
We only consider planar Wilson Loops: $C_s$ is a rectangle of temporal extent $T$ and spatial extent $R$. The quark-quark potential is given by $V(R)=\lim_{T\rightarrow\infty} \log(\langle W_L[C_s]\rangle)/T$.

To circumvent the path ordering, we split $C_s$ into infinitesimal segments of length $\delta \sigma$, and define the gauge link as $U_\sigma \in SU(N_C) = P[e^{-ig\int_{\sigma}^{\sigma + \delta \sigma} A_\sigma d\sigma}] \sim e^{-ig \delta \sigma A_\sigma}$. $0\le\sigma\le L$ represents the position along the curve and we write $A_\sigma \equiv A_{\mu(\sigma)}(x(\sigma))$. We have assumed that the gauge field is differentiable. $W[C_s]$ can then be written as
\begin{gather}
W[C_s] = \lim_{\delta \sigma \rightarrow 0} \prod_{\sigma = 0,\delta \sigma,2\delta\sigma,\ldots}^{L-\delta\sigma} U_\sigma.
\end{gather}

We proceed by inserting an identity operator between each pair of gauge links along the curve. The goal is to replace $U$ with an Abelian field, and evaluate the Wilson Loop using Stokes' theorem. Previous work~\cite{Kondo:2008su,*Shibata:2009af,*Kondo:2010pt,*Kondo:2005eq} has used the identity
$I = \int d\Theta \Theta \ket{e_{N_C}}\bra{e_{N_C}} \Theta^\dagger$, for $\Theta \in SU(N_C)/U(N_C-1)$ and $e_{N_C}$ a unit colour vector; but we choose differently, with no need for the integral and with $\Theta$ in a larger group so that it can `see' more types of monopole.

We introduce a field $\theta_\sigma \equiv \theta(x(\sigma))$, which, for the moment, we shall take to be an element of U($N_C$), at each point along $C_s$ and insert the identity operator $\theta_\sigma \theta_\sigma^\dagger$ between each of the gauge links. $\theta$ is chosen so that $\theta^\dagger_\sigma U_\sigma \theta_{\sigma + \delta\sigma}$ is diagonal. $\theta_s$ therefore contains the eigenvectors of $W[C_s]$: $W[C_s]\theta_s = \theta_s D$, where $D$ is some diagonal element of $SU(N_C)$. As the phases of the eigenvectors are arbitrary, this definition only determines $\theta$ up to a $U(N_C)$ transformation $\theta \rightarrow \theta \chi$. Fixing these phases and the ordering of the eigenvalues by some arbitrary \textit{fixing condition} gives a unique choice of $\theta \in SU(N_C)/(U(1))^{N_C-1}$. Under a gauge transformation $U_\sigma \rightarrow \Lambda_\sigma U_{\sigma}\Lambda_{\sigma + \delta\sigma}^\dagger$ for $\Lambda \in SU(N_C)$, $\theta \rightarrow \Lambda \theta \chi$,
where the $U(N_C)$ factor $\chi$ depends on the fixing condition. With $\theta^\dagger_\sigma U_\sigma \theta_{\sigma + \delta\sigma} = e^{i\sum_{\lambda^j \text{ diagonal}}  u_j \lambda^j}$,
\begin{gather}
\theta^\dagger_s W[C_s]\theta_s = e^{i \sum_{\lambda^j \text{ diagonal}} \lambda^j \oint_{C_s} dx_\mu u_j},
\end{gather}
removing the non-Abelian structure and the path ordering.

We may extend this definition of $\theta$ across all space by constructing nested curves in the same plane as $C_s$ and then stacking these curves on top of each other in the other dimensions. We then define $\theta$ so it diagonalises $W$ along each of these curves. We now introduce a restricted SU($N_C$) gauge field $\hat{U}_\mu$ which is diagonalised by $\theta$, so, for $\lambda^j$ a diagonal Gell-Mann matrix
\begin{align}
[\lambda^j,\theta_x^\dagger \hat{U}_{\mu,x}\theta_{x + \hat{\mu}\delta \sigma}] = &0 \nonumber\\
\hat{U}_{\mu,x} n^j_{x+\delta\sigma\hat{\mu} }\hat{U}^\dagger_{\mu,x} - n^j_x= &0&n_x^j \equiv &\theta_x \lambda^j \theta^\dagger_x,\label{eq:defeq1}
\end{align}
is satisfied across all of space-time.
We can introduce a second field $\hat{X}$ such that $U_\mu(x) = \hat{X}_\mu \hat{U}_\mu$, and restrict $\hat{X}_\mu$ by imposing the condition
\begin{align}
\tr [n^j_x(\hat{X}^\dagger_{\mu,x} - \hat{X}_{\mu,x})] = &0& \lambda^j&\text{ diagonal}\label{eq:defeq2}.
\end{align}
If there are multiple solutions to equations (\ref{eq:defeq1}) and (\ref{eq:defeq2}) we select the solution which maximises $\tr( X)$. Under a gauge transformation, $n_x \rightarrow \Lambda_x n_x \Lambda^\dagger_x$, $\hat{U}_\mu(x) \rightarrow \Lambda_x \hat{U}_{\mu,x} \Lambda^\dagger_{x+\hat{\mu}\delta\sigma}$ and $\hat{X}_{\mu,x} \rightarrow \Lambda_x \hat{X}_{\mu,x} \Lambda^\dagger_x$, so equations (\ref{eq:defeq1}) and (\ref{eq:defeq2}) are gauge invariant.
Equations (\ref{eq:defeq1}) and (\ref{eq:defeq2}) are the lattice versions of the defining equations of the CDG decomposition\cite{Cho:1980,*Cho:1981,*F-N:98,*Shabanov:1999,*Duan:1979}, which in the continuum is described by
\begin{align}
A_\mu = &\hat{A}_\mu + X_\mu&
D_\mu[\hat{A}] n^j = & 0\nonumber\\
0=&\tr(n^j X) &
D_\mu[\hat{A}] \alpha \equiv& \partial_\mu \alpha - i g [\hat{A},\alpha]\nonumber\\
\hat{A}_\mu =& \frac{1}{2}\sum_{\lambda_j\text{ diagonal}}\left[n^j\tr(n_j A_\mu) - \frac{1}{2} i g^{-1} [n^j,\partial_\mu n^j]\right].
\end{align}
The corresponding field strength is $\hat{F}_{\mu\nu}[\hatA] = n^j\hat{F}^j_{\mu\nu}$ with the gauge invariant\\
$
\hat{F}^j_{\mu\nu} = \frac{1}{2}\left[\partial_\mu\tr(n^j A_\nu) - \partial_\nu\tr(n^j A_\mu)\right] - \frac{1}{8}\tr (n^j [\partial_\mu n^k,\partial_\nu n^k]).
$
We express the restricted field as $\hat{U}_{\mu,x} \equiv \theta_{x}e^{i \sum_{\lambda_j\text{ diagonal}} \lambda_j \hat{u}^j_{\mu,x}} \theta^\dagger_{x+\hat{\mu}\delta\sigma}$, and since $\hat{U} = U$ along the curve $C_s$, we see that $W[C_s,U] = W[C_s,\hat{U}]=\theta_s W[C_s,\theta^\dagger \hat{U}\theta]\theta^\dagger_s$. Applying Stokes' theorem to the Abelian field $\theta^\dagger_{x} \hat{U}_{\mu,x} \theta_{x+\hat{\mu}\delta\sigma}$ gives
\begin{gather}
\theta^\dagger_s W[C_s]\theta_s = e^{i \sum_{\lambda_j \text{ diagonal}} \lambda_j \int dS_{\mu\nu} \hat{F}^j_{\mu\nu}},
\end{gather}
where $dS_{\mu\nu}$ is an element of the planar surface bounded by $C$.
Thus we may expect structures in the field strength $\hat{F}^j_{\mu\nu}$ to lead to a confining potential.

Whenever $\hat{U}$ is differentiable, $\hat{F}^j_{\mu\nu}$ is an exact derivative: $ \hat{F}^j_{\mu\nu}= \partial_\mu(\hat{u}^j_\nu) - \partial_\nu(\hat{u}^j_\mu)$~\cite{CundyForthcoming}. An area law scaling of the Wilson loop requires that $\hat{U}$ (and thus $\theta$) is non-differentiable at certain points. 
\section{Parametrisation and divergences of the $\theta$ field}\label{sec:3}
$\theta_s$ is defined as the matrix of eigenvectors of $W[C_s]$. It is discontinuous when $W[C_s]$ is discontinuous or has degenerate eigenvalues, and in another situation: in SU(2), we parametrise $\theta$ as
\begin{gather}
\theta = \left(\begin{array}{cc} \cos a& i\sin a e^{i c}\\i\sin a e^{-ic} & \cos a\end{array}\right)e^{id_3\lambda^3},
\end{gather}
with $c \in \mathbb{R}$, $0 \le a \le \pi/2$ and $d_3$ determined by the fixing condition. At both $a = 0$ and $a = \pi/2$, the parameter $c$ is ill-defined, and points where $a=\pi/2$ can lead to monopoles.
The parameter $c$ may wind itself around these points, creating a discontinuity in $\theta$. In the plane of the Wilson Loop, we parametrise space-time in polar coordinates $(r,\phi)$, with the origin at the point where $a = \pi/2$.
At some infinitesimal radius $r$, continuity of $\theta$ demands that $c(r,\phi =0) = c(r,\phi = 2\pi) + 2\pi \nu$ for integer winding number $\nu$, and if $c$ is ill-defined at $r = 0$ we may find $\nu \neq 0$, which implies a discontinuity in $\theta$.
We may evaluate $\hat{F}_{\mu\nu}$ using an integral of $\hat{u}$ around a loop of infinitesimal $r$ centred at $r=0$: $\hat{F}^j_{\mu\nu} = \delta(x)\oint d\tilde{\sigma} \hat{u}^j_{\tilde{\sigma}}$, and it can be shown that $\hat{u}_{\tilde{\sigma}}$ is proportional to $\partial_{\tilde{\sigma}} c$~\cite{CundyForthcoming}. This gives 
a $\delta$-function in $\hat{F}^j_{\mu\nu}$: a CFN magnetic monopole. It is reasonable to expect that the number of these monopoles will be proportional to the area of the Wilson Loop, leading to an area law scaling.


In SU(3), we parametrise $\theta$ in terms of six variables $0\le a_1\le\pi/2$, $0\le a_2\le \pi/2$, $0\le a_3\le \pi/2$, $c_1\in \mathbb{R}$, $c_2\in \mathbb{R}$, and $c_3\in \mathbb{R}$,  and two fixed constants, $d_3$ and $d_8$, as

\begin{gather}
{\tiny\theta \equiv\left(\begin{array}{ccc} 1&0&0\\0&\cos a_3& i\sin a_3 e^{i c_3}\\0&i\sin a_3 e^{-ic_3} & \cos a_3\end{array}\right)\left(\begin{array}{ccc} \cos a_2&0& i\sin a_2 e^{-i c_2}\\0&1&0\\i\sin a_2 e^{ic_2}&0 & \cos a_2\end{array}\right)\left(\begin{array}{ccc} \cos a_1& i\sin a_1 e^{i c_1}&0\\i\sin a_1 e^{-ic_1} & \cos a_1&0\\0&0&1\end{array}\right)e^{id_3\lambda^3 + i d^8 \lambda^8}.}
\end{gather}
There will be monopoles when one of the $a_i$ is $\pi/2$ and one of the $c_i$ winds itself around that point.

The area law scaling of $\langle \tr(W_L[C])\rangle$ may thus be related to $\delta$-functions in the reduced field strength caused by discontinuities in the $\theta$-field at $a=\pi/2$ (seen in the magnetic, $\tr (n[\partial_\mu n,\partial_\nu n) ]$, term within $\hat{F}_{\mu\nu}$) and characterised by a non-zero winding of the parameter $c$ around these monopoles.
\section{Numerical results}\label{sec:4}
We generated $16^332$ and $20^340$ quenched configurations in SU(3) with a Tadpole Improved Luscher-Weisz gauge action~\cite{TILW} using a Hybrid Monte Carlo routine~\cite{HMC} (see table \ref{tab:1}). The lattice spacing was measured using the string tension $\sigma \sim (420 \text{MeV})^2$. We fixed to the Landau gauge and applied ten steps of improved stout smearing~\cite{Morningstar:2003gk,*Moran:2008ra} with parameters $\rho = 0.015$ and $\epsilon = 0$. $\theta$ and $\hat{U}$ were calculated numerically.
\begin{table}
{\tiny
\begin{center}
\begin{tabular}{|l l l l l|}
\hline
Lattice size (lattice units)& Spatial Lattice size (fm)&$\beta$&lattice spacing (fm)& Number of configurations\\
\hline
$16^332$& 2.30& 8.0&0.144(1) &91
\\
$16^332$&1.84& 8.3 &0.115(1) &91
\\
$16^332$&1.58 & 8.52 &0.099(1) &82
\\
$20^340$&2.30 & 8.3&0.115(1) &20\\
\hline
\end{tabular}
\end{center}
}
\caption{Parameters for our simulations. }\label{tab:1}
\end{table}

\begin{figure}
\begin{center}
\begin{tabular}{c}
{\tiny
{
\begingroup
  \makeatletter
  \providecommand\color[2][]{%
    \GenericError{(gnuplot) \space\space\space\@spaces}{%
      Package color not loaded in conjunction with
      terminal option `colourtext'%
    }{See the gnuplot documentation for explanation.%
    }{Either use 'blacktext' in gnuplot or load the package
      color.sty in LaTeX.}%
    \renewcommand\color[2][]{}%
  }%
  \providecommand\includegraphics[2][]{%
    \GenericError{(gnuplot) \space\space\space\@spaces}{%
      Package graphicx or graphics not loaded%
    }{See the gnuplot documentation for explanation.%
    }{The gnuplot epslatex terminal needs graphicx.sty or graphics.sty.}%
    \renewcommand\includegraphics[2][]{}%
  }%
  \providecommand\rotatebox[2]{#2}%
  \@ifundefined{ifGPcolor}{%
    \newif\ifGPcolor
    \GPcolortrue
  }{}%
  \@ifundefined{ifGPblacktext}{%
    \newif\ifGPblacktext
    \GPblacktexttrue
  }{}%
  \let\gplgaddtomacro\g@addto@macro
  \gdef\gplbacktext{}%
  \gdef\gplfronttext{}%
  \makeatother
  \ifGPblacktext
    \def\colorrgb#1{}%
    \def\colorgray#1{}%
  \else
    \ifGPcolor
      \def\colorrgb#1{\color[rgb]{#1}}%
      \def\colorgray#1{\color[gray]{#1}}%
      \expandafter\def\csname LTw\endcsname{\color{white}}%
      \expandafter\def\csname LTb\endcsname{\color{black}}%
      \expandafter\def\csname LTa\endcsname{\color{black}}%
      \expandafter\def\csname LT0\endcsname{\color[rgb]{1,0,0}}%
      \expandafter\def\csname LT1\endcsname{\color[rgb]{0,1,0}}%
      \expandafter\def\csname LT2\endcsname{\color[rgb]{0,0,1}}%
      \expandafter\def\csname LT3\endcsname{\color[rgb]{1,0,1}}%
      \expandafter\def\csname LT4\endcsname{\color[rgb]{0,1,1}}%
      \expandafter\def\csname LT5\endcsname{\color[rgb]{1,1,0}}%
      \expandafter\def\csname LT6\endcsname{\color[rgb]{0,0,0}}%
      \expandafter\def\csname LT7\endcsname{\color[rgb]{1,0.3,0}}%
      \expandafter\def\csname LT8\endcsname{\color[rgb]{0.5,0.5,0.5}}%
    \else
      \def\colorrgb#1{\color{black}}%
      \def\colorgray#1{\color[gray]{#1}}%
      \expandafter\def\csname LTw\endcsname{\color{white}}%
      \expandafter\def\csname LTb\endcsname{\color{black}}%
      \expandafter\def\csname LTa\endcsname{\color{black}}%
      \expandafter\def\csname LT0\endcsname{\color{black}}%
      \expandafter\def\csname LT1\endcsname{\color{black}}%
      \expandafter\def\csname LT2\endcsname{\color{black}}%
      \expandafter\def\csname LT3\endcsname{\color{black}}%
      \expandafter\def\csname LT4\endcsname{\color{black}}%
      \expandafter\def\csname LT5\endcsname{\color{black}}%
      \expandafter\def\csname LT6\endcsname{\color{black}}%
      \expandafter\def\csname LT7\endcsname{\color{black}}%
      \expandafter\def\csname LT8\endcsname{\color{black}}%
    \fi
  \fi
  \setlength{\unitlength}{0.0500bp}%
  \begin{picture}(7370.00,2692.00)%
    \gplgaddtomacro\gplbacktext{%
      \csname LTb\endcsname%
      \put(740,779){\makebox(0,0)[r]{\strut{} 2}}%
      \put(740,1058){\makebox(0,0)[r]{\strut{} 4}}%
      \put(740,1337){\makebox(0,0)[r]{\strut{} 6}}%
      \put(740,1615){\makebox(0,0)[r]{\strut{} 8}}%
      \put(740,1894){\makebox(0,0)[r]{\strut{} 10}}%
      \put(740,2172){\makebox(0,0)[r]{\strut{} 12}}%
      \put(740,2451){\makebox(0,0)[r]{\strut{} 14}}%
      \put(1572,440){\makebox(0,0){\strut{} 5}}%
      \put(2462,440){\makebox(0,0){\strut{} 10}}%
      \put(3352,440){\makebox(0,0){\strut{} 15}}%
      \put(4242,440){\makebox(0,0){\strut{} 20}}%
      \put(5132,440){\makebox(0,0){\strut{} 25}}%
      \put(6022,440){\makebox(0,0){\strut{} 30}}%
      \put(160,1545){\rotatebox{-270}{\makebox(0,0){\strut{}X}}}%
      \put(3441,140){\makebox(0,0){\strut{}T}}%
    }%
    \gplgaddtomacro\gplfronttext{%
      \csname LTb\endcsname%
      \put(6529,821){\makebox(0,0)[l]{\strut{}-2}}%
      \put(6529,1183){\makebox(0,0)[l]{\strut{}-1}}%
      \put(6529,1545){\makebox(0,0)[l]{\strut{} 0}}%
      \put(6529,1907){\makebox(0,0)[l]{\strut{} 1}}%
      \put(6529,2269){\makebox(0,0)[l]{\strut{} 2}}%
      \put(6829,1545){\rotatebox{-270}{\makebox(0,0){\strut{}\rotatebox{-90}{$\hat{F}_{xt}$}}}}%
    }%
    \gplbacktext
    \put(0,0){\includegraphics{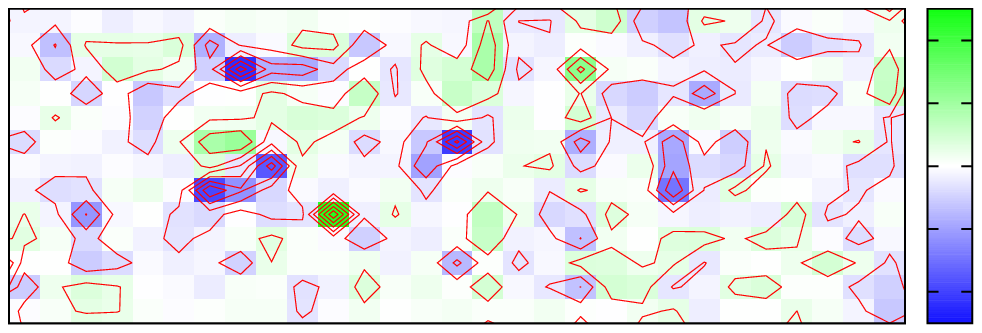}}%
    \gplfronttext
  \end{picture}%
\endgroup

}
}
\\
{\tiny
\begingroup
  \makeatletter
  \providecommand\color[2][]{%
    \GenericError{(gnuplot) \space\space\space\@spaces}{%
      Package color not loaded in conjunction with
      terminal option `colourtext'%
    }{See the gnuplot documentation for explanation.%
    }{Either use 'blacktext' in gnuplot or load the package
      color.sty in LaTeX.}%
    \renewcommand\color[2][]{}%
  }%
  \providecommand\includegraphics[2][]{%
    \GenericError{(gnuplot) \space\space\space\@spaces}{%
      Package graphicx or graphics not loaded%
    }{See the gnuplot documentation for explanation.%
    }{The gnuplot epslatex terminal needs graphicx.sty or graphics.sty.}%
    \renewcommand\includegraphics[2][]{}%
  }%
  \providecommand\rotatebox[2]{#2}%
  \@ifundefined{ifGPcolor}{%
    \newif\ifGPcolor
    \GPcolortrue
  }{}%
  \@ifundefined{ifGPblacktext}{%
    \newif\ifGPblacktext
    \GPblacktexttrue
  }{}%
  \let\gplgaddtomacro\g@addto@macro
  \gdef\gplbacktext{}%
  \gdef\gplfronttext{}%
  \makeatother
  \ifGPblacktext
    \def\colorrgb#1{}%
    \def\colorgray#1{}%
  \else
    \ifGPcolor
      \def\colorrgb#1{\color[rgb]{#1}}%
      \def\colorgray#1{\color[gray]{#1}}%
      \expandafter\def\csname LTw\endcsname{\color{white}}%
      \expandafter\def\csname LTb\endcsname{\color{black}}%
      \expandafter\def\csname LTa\endcsname{\color{black}}%
      \expandafter\def\csname LT0\endcsname{\color[rgb]{1,0,0}}%
      \expandafter\def\csname LT1\endcsname{\color[rgb]{0,1,0}}%
      \expandafter\def\csname LT2\endcsname{\color[rgb]{0,0,1}}%
      \expandafter\def\csname LT3\endcsname{\color[rgb]{1,0,1}}%
      \expandafter\def\csname LT4\endcsname{\color[rgb]{0,1,1}}%
      \expandafter\def\csname LT5\endcsname{\color[rgb]{1,1,0}}%
      \expandafter\def\csname LT6\endcsname{\color[rgb]{0,0,0}}%
      \expandafter\def\csname LT7\endcsname{\color[rgb]{1,0.3,0}}%
      \expandafter\def\csname LT8\endcsname{\color[rgb]{0.5,0.5,0.5}}%
    \else
      \def\colorrgb#1{\color{black}}%
      \def\colorgray#1{\color[gray]{#1}}%
      \expandafter\def\csname LTw\endcsname{\color{white}}%
      \expandafter\def\csname LTb\endcsname{\color{black}}%
      \expandafter\def\csname LTa\endcsname{\color{black}}%
      \expandafter\def\csname LT0\endcsname{\color{black}}%
      \expandafter\def\csname LT1\endcsname{\color{black}}%
      \expandafter\def\csname LT2\endcsname{\color{black}}%
      \expandafter\def\csname LT3\endcsname{\color{black}}%
      \expandafter\def\csname LT4\endcsname{\color{black}}%
      \expandafter\def\csname LT5\endcsname{\color{black}}%
      \expandafter\def\csname LT6\endcsname{\color{black}}%
      \expandafter\def\csname LT7\endcsname{\color{black}}%
      \expandafter\def\csname LT8\endcsname{\color{black}}%
    \fi
  \fi
  \setlength{\unitlength}{0.0500bp}%
  \begin{picture}(7370.00,2834.00)%
    \gplgaddtomacro\gplbacktext{%
      \csname LTb\endcsname%
      \put(860,640){\makebox(0,0)[r]{\strut{} 0}}%
      \put(860,817){\makebox(0,0)[r]{\strut{} 0.2}}%
      \put(860,994){\makebox(0,0)[r]{\strut{} 0.4}}%
      \put(860,1171){\makebox(0,0)[r]{\strut{} 0.6}}%
      \put(860,1348){\makebox(0,0)[r]{\strut{} 0.8}}%
      \put(860,1525){\makebox(0,0)[r]{\strut{} 1}}%
      \put(860,1702){\makebox(0,0)[r]{\strut{} 1.2}}%
      \put(860,1879){\makebox(0,0)[r]{\strut{} 1.4}}%
      \put(860,2056){\makebox(0,0)[r]{\strut{} 1.6}}%
      \put(860,2233){\makebox(0,0)[r]{\strut{} 1.8}}%
      \put(1587,440){\makebox(0,0){\strut{} 2}}%
      \put(2396,440){\makebox(0,0){\strut{} 4}}%
      \put(3205,440){\makebox(0,0){\strut{} 6}}%
      \put(4015,440){\makebox(0,0){\strut{} 8}}%
      \put(4824,440){\makebox(0,0){\strut{} 10}}%
      \put(160,1436){\rotatebox{-270}{\makebox(0,0){\strut{}$\log(\langle W[R,T] \rangle )/T$}}}%
      \put(3003,140){\makebox(0,0){\strut{}R}}%
      \put(3003,2533){\makebox(0,0){\strut{}$\beta = 8.52$ Peaks in $\hat{F}_{xt}$ excluded, $T = 6$}}%
    }%
    \gplgaddtomacro\gplfronttext{%
      \csname LTb\endcsname%
      \put(6466,2133){\makebox(0,0)[r]{\strut{}$\mathcal{C}=0.990$}}%
      \csname LTb\endcsname%
      \put(6466,1933){\makebox(0,0)[r]{\strut{}$\mathcal{C}=0.505$}}%
      \csname LTb\endcsname%
      \put(6466,1733){\makebox(0,0)[r]{\strut{}$\mathcal{C} = 0.256$}}%
      \csname LTb\endcsname%
      \put(6466,1533){\makebox(0,0)[r]{\strut{}$\mathcal{C}=0.131$}}%
      \csname LTb\endcsname%
      \put(6466,1333){\makebox(0,0)[r]{\strut{}$\mathcal{C}=0.067$}}%
    }%
    \gplbacktext
    \put(0,0){\includegraphics{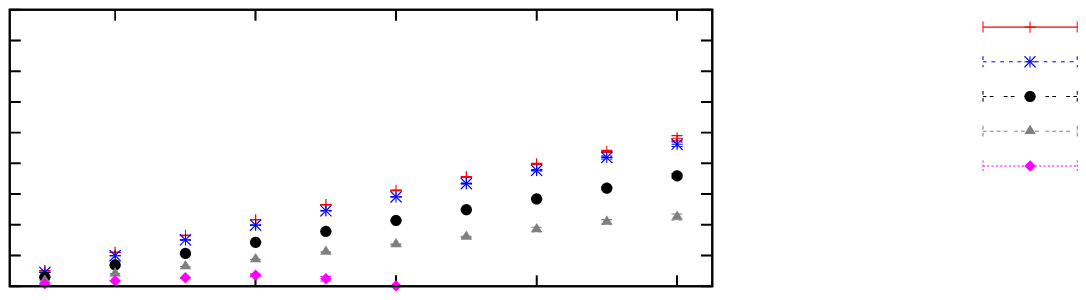}}%
    \gplfronttext
  \end{picture}%
\endgroup

}
\end{tabular}
\end{center}
\caption{A cross section of the field strength $\hat{F}_{xt}^3$ at $z = 0, y=10$ on one configuration (top), and the $\hat{U}$ string tension excluding peaks of height $|F_{xt}| > \mathcal{C} \hat{F}^{(max)}$ from the average over Wilson loops (bottom). Due to the limited resolution of the lattice, the extrapolated contour lines have an error of up to one lattice spacing.}\label{fig:1}
\end{figure}

In figure \ref{fig:1} (top plot), we plot a slice of the restricted field strength at fixed $y$ and $z$ coordinates, and see that it is indeed dominated by peaks a few lattice spacings across. Plots on neighbouring slices of the lattice show a very different pattern, suggesting that these are point-like objects rather than strings or membranes. In figure \ref{fig:1} (bottom), we show that these peaks are responsible for the string tension by plotting the quark potential only averaging over those Wilson loops in the $xt$ plane which exclude these peaks: only including loops where $|F_{xt}(x)| < \mathcal{C} F^{(max)}_{xt}(x)$, $\forall x$ within the curve $C_s$. $F^{(max)}$ is the configuration maximum value of the field strength. The string tension decreases as more of these objects are excluded, suggesting that these maxima indeed cause the area law scaling.

\begin{figure}
\begin{center}
\begin{tabular}{c}
{\tiny
\begingroup
  \makeatletter
  \providecommand\color[2][]{%
    \GenericError{(gnuplot) \space\space\space\@spaces}{%
      Package color not loaded in conjunction with
      terminal option `colourtext'%
    }{See the gnuplot documentation for explanation.%
    }{Either use 'blacktext' in gnuplot or load the package
      color.sty in LaTeX.}%
    \renewcommand\color[2][]{}%
  }%
  \providecommand\includegraphics[2][]{%
    \GenericError{(gnuplot) \space\space\space\@spaces}{%
      Package graphicx or graphics not loaded%
    }{See the gnuplot documentation for explanation.%
    }{The gnuplot epslatex terminal needs graphicx.sty or graphics.sty.}%
    \renewcommand\includegraphics[2][]{}%
  }%
  \providecommand\rotatebox[2]{#2}%
  \@ifundefined{ifGPcolor}{%
    \newif\ifGPcolor
    \GPcolortrue
  }{}%
  \@ifundefined{ifGPblacktext}{%
    \newif\ifGPblacktext
    \GPblacktexttrue
  }{}%
  \let\gplgaddtomacro\g@addto@macro
  \gdef\gplbacktext{}%
  \gdef\gplfronttext{}%
  \makeatother
  \ifGPblacktext
    \def\colorrgb#1{}%
    \def\colorgray#1{}%
  \else
    \ifGPcolor
      \def\colorrgb#1{\color[rgb]{#1}}%
      \def\colorgray#1{\color[gray]{#1}}%
      \expandafter\def\csname LTw\endcsname{\color{white}}%
      \expandafter\def\csname LTb\endcsname{\color{black}}%
      \expandafter\def\csname LTa\endcsname{\color{black}}%
      \expandafter\def\csname LT0\endcsname{\color[rgb]{1,0,0}}%
      \expandafter\def\csname LT1\endcsname{\color[rgb]{0,1,0}}%
      \expandafter\def\csname LT2\endcsname{\color[rgb]{0,0,1}}%
      \expandafter\def\csname LT3\endcsname{\color[rgb]{1,0,1}}%
      \expandafter\def\csname LT4\endcsname{\color[rgb]{0,1,1}}%
      \expandafter\def\csname LT5\endcsname{\color[rgb]{1,1,0}}%
      \expandafter\def\csname LT6\endcsname{\color[rgb]{0,0,0}}%
      \expandafter\def\csname LT7\endcsname{\color[rgb]{1,0.3,0}}%
      \expandafter\def\csname LT8\endcsname{\color[rgb]{0.5,0.5,0.5}}%
    \else
      \def\colorrgb#1{\color{black}}%
      \def\colorgray#1{\color[gray]{#1}}%
      \expandafter\def\csname LTw\endcsname{\color{white}}%
      \expandafter\def\csname LTb\endcsname{\color{black}}%
      \expandafter\def\csname LTa\endcsname{\color{black}}%
      \expandafter\def\csname LT0\endcsname{\color{black}}%
      \expandafter\def\csname LT1\endcsname{\color{black}}%
      \expandafter\def\csname LT2\endcsname{\color{black}}%
      \expandafter\def\csname LT3\endcsname{\color{black}}%
      \expandafter\def\csname LT4\endcsname{\color{black}}%
      \expandafter\def\csname LT5\endcsname{\color{black}}%
      \expandafter\def\csname LT6\endcsname{\color{black}}%
      \expandafter\def\csname LT7\endcsname{\color{black}}%
      \expandafter\def\csname LT8\endcsname{\color{black}}%
    \fi
  \fi
  \setlength{\unitlength}{0.0500bp}%
  \begin{picture}(7370.00,2692.00)%
    \gplgaddtomacro\gplbacktext{%
      \csname LTb\endcsname%
      \put(740,640){\makebox(0,0)[r]{\strut{} 4}}%
      \put(740,866){\makebox(0,0)[r]{\strut{} 5}}%
      \put(740,1093){\makebox(0,0)[r]{\strut{} 6}}%
      \put(740,1319){\makebox(0,0)[r]{\strut{} 7}}%
      \put(740,1546){\makebox(0,0)[r]{\strut{} 8}}%
      \put(740,1772){\makebox(0,0)[r]{\strut{} 9}}%
      \put(740,1998){\makebox(0,0)[r]{\strut{} 10}}%
      \put(740,2225){\makebox(0,0)[r]{\strut{} 11}}%
      \put(740,2451){\makebox(0,0)[r]{\strut{} 12}}%
      \put(1075,440){\makebox(0,0){\strut{} 5}}%
      \put(2151,440){\makebox(0,0){\strut{} 10}}%
      \put(3226,440){\makebox(0,0){\strut{} 15}}%
      \put(4301,440){\makebox(0,0){\strut{} 20}}%
      \put(5377,440){\makebox(0,0){\strut{} 25}}%
      \put(160,1545){\rotatebox{-270}{\makebox(0,0){\strut{}X}}}%
      \put(3441,140){\makebox(0,0){\strut{}T}}%
    }%
    \gplgaddtomacro\gplfronttext{%
      \csname LTb\endcsname%
      \put(6529,640){\makebox(0,0)[l]{\strut{} 0}}%
      \put(6529,1002){\makebox(0,0)[l]{\strut{} 0.2}}%
      \put(6529,1364){\makebox(0,0)[l]{\strut{} 0.4}}%
      \put(6529,1726){\makebox(0,0)[l]{\strut{} 0.6}}%
      \put(6529,2088){\makebox(0,0)[l]{\strut{} 0.8}}%
      \put(6529,2451){\makebox(0,0)[l]{\strut{} 1}}%
      \put(7069,1545){\rotatebox{-270}{\makebox(0,0){\strut{}\rotatebox{-90}{$\frac{2 a}{\pi}$}}}}%
    }%
    \gplbacktext
    \put(0,0){\includegraphics{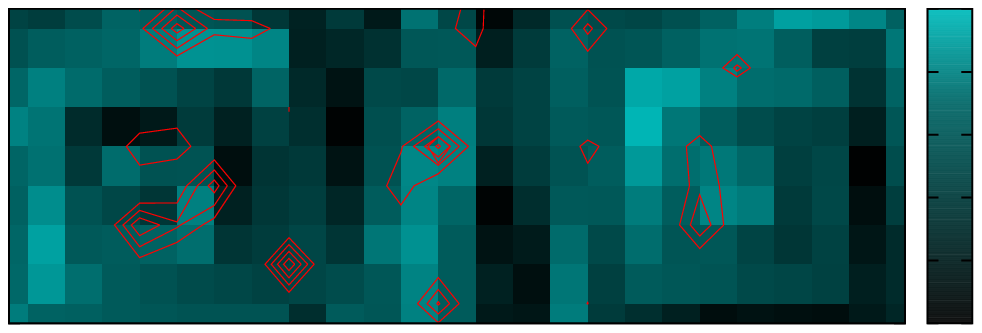}}%
    \gplfronttext
  \end{picture}%
\endgroup

}\\
{\tiny
\begingroup
  \makeatletter
  \providecommand\color[2][]{%
    \GenericError{(gnuplot) \space\space\space\@spaces}{%
      Package color not loaded in conjunction with
      terminal option `colourtext'%
    }{See the gnuplot documentation for explanation.%
    }{Either use 'blacktext' in gnuplot or load the package
      color.sty in LaTeX.}%
    \renewcommand\color[2][]{}%
  }%
  \providecommand\includegraphics[2][]{%
    \GenericError{(gnuplot) \space\space\space\@spaces}{%
      Package graphicx or graphics not loaded%
    }{See the gnuplot documentation for explanation.%
    }{The gnuplot epslatex terminal needs graphicx.sty or graphics.sty.}%
    \renewcommand\includegraphics[2][]{}%
  }%
  \providecommand\rotatebox[2]{#2}%
  \@ifundefined{ifGPcolor}{%
    \newif\ifGPcolor
    \GPcolortrue
  }{}%
  \@ifundefined{ifGPblacktext}{%
    \newif\ifGPblacktext
    \GPblacktexttrue
  }{}%
  \let\gplgaddtomacro\g@addto@macro
  \gdef\gplbacktext{}%
  \gdef\gplfronttext{}%
  \makeatother
  \ifGPblacktext
    \def\colorrgb#1{}%
    \def\colorgray#1{}%
  \else
    \ifGPcolor
      \def\colorrgb#1{\color[rgb]{#1}}%
      \def\colorgray#1{\color[gray]{#1}}%
      \expandafter\def\csname LTw\endcsname{\color{white}}%
      \expandafter\def\csname LTb\endcsname{\color{black}}%
      \expandafter\def\csname LTa\endcsname{\color{black}}%
      \expandafter\def\csname LT0\endcsname{\color[rgb]{1,0,0}}%
      \expandafter\def\csname LT1\endcsname{\color[rgb]{0,1,0}}%
      \expandafter\def\csname LT2\endcsname{\color[rgb]{0,0,1}}%
      \expandafter\def\csname LT3\endcsname{\color[rgb]{1,0,1}}%
      \expandafter\def\csname LT4\endcsname{\color[rgb]{0,1,1}}%
      \expandafter\def\csname LT5\endcsname{\color[rgb]{1,1,0}}%
      \expandafter\def\csname LT6\endcsname{\color[rgb]{0,0,0}}%
      \expandafter\def\csname LT7\endcsname{\color[rgb]{1,0.3,0}}%
      \expandafter\def\csname LT8\endcsname{\color[rgb]{0.5,0.5,0.5}}%
    \else
      \def\colorrgb#1{\color{black}}%
      \def\colorgray#1{\color[gray]{#1}}%
      \expandafter\def\csname LTw\endcsname{\color{white}}%
      \expandafter\def\csname LTb\endcsname{\color{black}}%
      \expandafter\def\csname LTa\endcsname{\color{black}}%
      \expandafter\def\csname LT0\endcsname{\color{black}}%
      \expandafter\def\csname LT1\endcsname{\color{black}}%
      \expandafter\def\csname LT2\endcsname{\color{black}}%
      \expandafter\def\csname LT3\endcsname{\color{black}}%
      \expandafter\def\csname LT4\endcsname{\color{black}}%
      \expandafter\def\csname LT5\endcsname{\color{black}}%
      \expandafter\def\csname LT6\endcsname{\color{black}}%
      \expandafter\def\csname LT7\endcsname{\color{black}}%
      \expandafter\def\csname LT8\endcsname{\color{black}}%
    \fi
  \fi
  \setlength{\unitlength}{0.0500bp}%
  \begin{picture}(7370.00,2692.00)%
    \gplgaddtomacro\gplbacktext{%
      \csname LTb\endcsname%
      \put(740,640){\makebox(0,0)[r]{\strut{} 4}}%
      \put(740,866){\makebox(0,0)[r]{\strut{} 5}}%
      \put(740,1093){\makebox(0,0)[r]{\strut{} 6}}%
      \put(740,1319){\makebox(0,0)[r]{\strut{} 7}}%
      \put(740,1546){\makebox(0,0)[r]{\strut{} 8}}%
      \put(740,1772){\makebox(0,0)[r]{\strut{} 9}}%
      \put(740,1998){\makebox(0,0)[r]{\strut{} 10}}%
      \put(740,2225){\makebox(0,0)[r]{\strut{} 11}}%
      \put(740,2451){\makebox(0,0)[r]{\strut{} 12}}%
      \put(1075,440){\makebox(0,0){\strut{} 5}}%
      \put(2151,440){\makebox(0,0){\strut{} 10}}%
      \put(3226,440){\makebox(0,0){\strut{} 15}}%
      \put(4301,440){\makebox(0,0){\strut{} 20}}%
      \put(5377,440){\makebox(0,0){\strut{} 25}}%
      \put(160,1545){\rotatebox{-270}{\makebox(0,0){\strut{}X}}}%
      \put(3441,140){\makebox(0,0){\strut{}T}}%
    }%
    \gplgaddtomacro\gplfronttext{%
      \csname LTb\endcsname%
      \put(6529,640){\makebox(0,0)[l]{\strut{} 0}}%
      \put(6529,927){\makebox(0,0)[l]{\strut{} 1}}%
      \put(6529,1214){\makebox(0,0)[l]{\strut{} 2}}%
      \put(6529,1502){\makebox(0,0)[l]{\strut{} 3}}%
      \put(6529,1789){\makebox(0,0)[l]{\strut{} 4}}%
      \put(6529,2077){\makebox(0,0)[l]{\strut{} 5}}%
      \put(6529,2364){\makebox(0,0)[l]{\strut{} 6}}%
      \put(6829,1545){\rotatebox{-270}{\makebox(0,0){\strut{}\rotatebox{-90}{$c$}}}}%
    }%
    \gplbacktext
    \put(0,0){\includegraphics{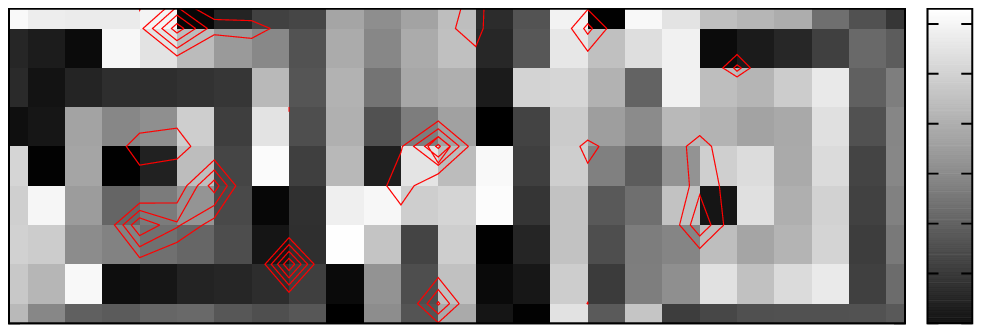}}%
    \gplfronttext
  \end{picture}%
\endgroup

}
\end{tabular}
\end{center}

\caption{A comparison between the peaks in $\hat{F}^{3,8}_{xt}$ (red contours) against the peaks in $2a_1/\pi$ (shaded background, top) and the angle $c_2$ (shaded background, bottom)}\label{fig:4}
\end{figure}

In figure \ref{fig:4} (top), we investigate whether the peaks in $F_{xt}$ are close to the maxima of $a_i$ (on the lattice, these will not be precisely at $a = \pi/2$). We have only shown data for one of the $a_i$ parameters; those peaks which are not close to a maximum of $a$ in this plot are near the maximum of a different $a_i$. The contours lines show the maxima of $\hat{F}_{xt}^j$, combining the two components of the gauge field, while the level of shading shows the values of $a_i$, with a dark colour indicating $a_i \sim 0$ and a light colour $a_i \sim \pi/2$. The overall picture is a little ambiguous, but consistent with expectations. We also show a similar plot for $c_i$ (figure \ref{fig:4} bottom). This should wind itself around the maxima, shown by a gradual darkening of the background as we rotate around the peak. Again, we only show data for one of
the three $c_i$ but we see some examples which suggest that $c$ indeed winds itself around the peak (for example at the center of the plot).

\begin{figure}
\begin{center}
\begin{tabular}{c c}
{\tiny
\begingroup
  \makeatletter
  \providecommand\color[2][]{%
    \GenericError{(gnuplot) \space\space\space\@spaces}{%
      Package color not loaded in conjunction with
      terminal option `colourtext'%
    }{See the gnuplot documentation for explanation.%
    }{Either use 'blacktext' in gnuplot or load the package
      color.sty in LaTeX.}%
    \renewcommand\color[2][]{}%
  }%
  \providecommand\includegraphics[2][]{%
    \GenericError{(gnuplot) \space\space\space\@spaces}{%
      Package graphicx or graphics not loaded%
    }{See the gnuplot documentation for explanation.%
    }{The gnuplot epslatex terminal needs graphicx.sty or graphics.sty.}%
    \renewcommand\includegraphics[2][]{}%
  }%
  \providecommand\rotatebox[2]{#2}%
  \@ifundefined{ifGPcolor}{%
    \newif\ifGPcolor
    \GPcolortrue
  }{}%
  \@ifundefined{ifGPblacktext}{%
    \newif\ifGPblacktext
    \GPblacktexttrue
  }{}%
  \let\gplgaddtomacro\g@addto@macro
  \gdef\gplbacktext{}%
  \gdef\gplfronttext{}%
  \makeatother
  \ifGPblacktext
    \def\colorrgb#1{}%
    \def\colorgray#1{}%
  \else
    \ifGPcolor
      \def\colorrgb#1{\color[rgb]{#1}}%
      \def\colorgray#1{\color[gray]{#1}}%
      \expandafter\def\csname LTw\endcsname{\color{white}}%
      \expandafter\def\csname LTb\endcsname{\color{black}}%
      \expandafter\def\csname LTa\endcsname{\color{black}}%
      \expandafter\def\csname LT0\endcsname{\color[rgb]{1,0,0}}%
      \expandafter\def\csname LT1\endcsname{\color[rgb]{0,1,0}}%
      \expandafter\def\csname LT2\endcsname{\color[rgb]{0,0,1}}%
      \expandafter\def\csname LT3\endcsname{\color[rgb]{1,0,1}}%
      \expandafter\def\csname LT4\endcsname{\color[rgb]{0,1,1}}%
      \expandafter\def\csname LT5\endcsname{\color[rgb]{1,1,0}}%
      \expandafter\def\csname LT6\endcsname{\color[rgb]{0,0,0}}%
      \expandafter\def\csname LT7\endcsname{\color[rgb]{1,0.3,0}}%
      \expandafter\def\csname LT8\endcsname{\color[rgb]{0.5,0.5,0.5}}%
    \else
      \def\colorrgb#1{\color{black}}%
      \def\colorgray#1{\color[gray]{#1}}%
      \expandafter\def\csname LTw\endcsname{\color{white}}%
      \expandafter\def\csname LTb\endcsname{\color{black}}%
      \expandafter\def\csname LTa\endcsname{\color{black}}%
      \expandafter\def\csname LT0\endcsname{\color{black}}%
      \expandafter\def\csname LT1\endcsname{\color{black}}%
      \expandafter\def\csname LT2\endcsname{\color{black}}%
      \expandafter\def\csname LT3\endcsname{\color{black}}%
      \expandafter\def\csname LT4\endcsname{\color{black}}%
      \expandafter\def\csname LT5\endcsname{\color{black}}%
      \expandafter\def\csname LT6\endcsname{\color{black}}%
      \expandafter\def\csname LT7\endcsname{\color{black}}%
      \expandafter\def\csname LT8\endcsname{\color{black}}%
    \fi
  \fi
  \setlength{\unitlength}{0.0500bp}%
  \begin{picture}(4818.00,2834.00)%
    \gplgaddtomacro\gplbacktext{%
      \csname LTb\endcsname%
      \put(860,640){\makebox(0,0)[r]{\strut{} 0}}%
      \put(860,817){\makebox(0,0)[r]{\strut{} 0.2}}%
      \put(860,994){\makebox(0,0)[r]{\strut{} 0.4}}%
      \put(860,1171){\makebox(0,0)[r]{\strut{} 0.6}}%
      \put(860,1348){\makebox(0,0)[r]{\strut{} 0.8}}%
      \put(860,1525){\makebox(0,0)[r]{\strut{} 1}}%
      \put(860,1702){\makebox(0,0)[r]{\strut{} 1.2}}%
      \put(860,1879){\makebox(0,0)[r]{\strut{} 1.4}}%
      \put(860,2056){\makebox(0,0)[r]{\strut{} 1.6}}%
      \put(860,2233){\makebox(0,0)[r]{\strut{} 1.8}}%
      \put(1502,440){\makebox(0,0){\strut{} 2}}%
      \put(2197,440){\makebox(0,0){\strut{} 4}}%
      \put(2892,440){\makebox(0,0){\strut{} 6}}%
      \put(3588,440){\makebox(0,0){\strut{} 8}}%
      \put(4283,440){\makebox(0,0){\strut{} 10}}%
      \put(160,1436){\rotatebox{-270}{\makebox(0,0){\strut{}$\log(\langle W[R,T]\rangle)/T$}}}%
      \put(2718,140){\makebox(0,0){\strut{}R}}%
      \put(2718,2533){\makebox(0,0){\strut{}$\beta = 8.3$}}%
    }%
    \gplgaddtomacro\gplfronttext{%
      \csname LTb\endcsname%
      \put(2109,2044){\makebox(0,0)[r]{\strut{}$U$ field $T = \infty$}}%
      \csname LTb\endcsname%
      \put(2109,1844){\makebox(0,0)[r]{\strut{}$\hat{U}$ field $T = \infty$}}%
      \csname LTb\endcsname%
      \put(2109,1644){\makebox(0,0)[r]{\strut{}$\tilde{U}$ field $T = \infty$}}%
      \csname LTb\endcsname%
      \put(2109,1444){\makebox(0,0)[r]{\strut{}$M$ field $T = \infty$}}%
    }%
    \gplbacktext
    \put(0,0){\includegraphics{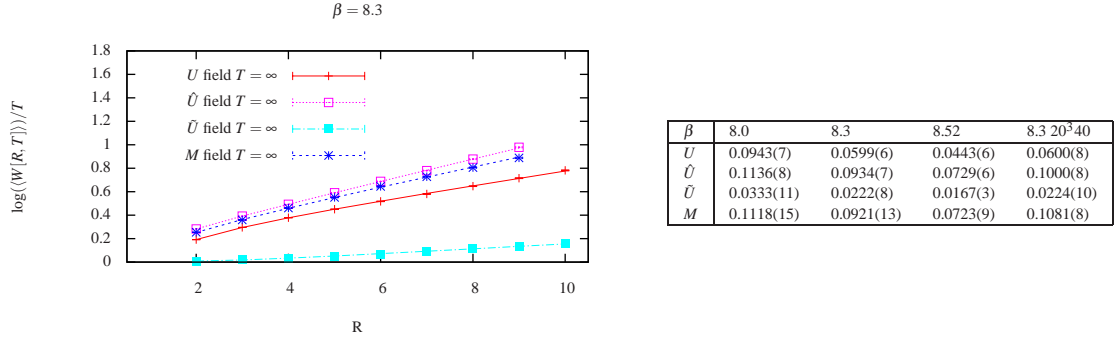}}%
    \gplfronttext
  \end{picture}%
\endgroup

}
&\raisebox{3\height}{\tiny\begin{tabular}{|l|l l l l|}
\hline
$\beta$&8.0&8.3&8.52&8.3 $20^340$\\
\hline
$U$&0.0943(7)&0.0599(6)&0.0443(6)&0.0600(8)
\\
$\hat{U}$&0.1136(8)&0.0934(7)&0.0729(6)&0.1000(8)
\\
$\tilde{U}$& 0.0333(11)&0.0222(8) & 0.0167(3)&0.0224(10)
\\
$M$ &0.1118(15)  &0.0921(13) & 0.0723(9)&0.1081(8)
\\ \hline
\end{tabular}
}
\end{tabular}
\end{center}
\caption{The string tension extrapolated to infinite time for the original gauge field $U$, the restricted gauge field $\hat{U}$, the over-smeared field $\tilde{U}$ and the monopole field $M$ (left); the fit results for the string tension (right)}\label{fig:5}
\end{figure}

In figure \ref{fig:5} we compare the string tension for the original gauge field $U$, the decomposed gauge field $\hat{U}$, and the monopole contribution.
The monopole string tension is extracted from the CDG decomposition of $\theta_{\mu,x} = \theta_x^\dagger \tilde{U}_{\mu,x} \theta_{x+ a \hat{\mu}}$ (rather than the non-gauge invariant $\theta_x^\dagger \theta_{x+ a \hat{\mu}}$ which gives too noisy data), with $\tilde{U}_{\mu,x}$ the gauge field subjected to 600 sweeps of stout-smearing at $\rho = 0.1$: enough smearing will destroy any structure, and
 any signal will be from the $\theta$ field calculated without smearing. The $\hat{U}$ field is identically equal to the $U$ field on those Wilson Loops used to define $\theta$; but to increase statistics, we averaged $\hat{U}$ over every Wilson Loop; so our results for the $\hat{U}$ and $U$ fields differ. Both the restricted field and the monopole field dominate the string tension.

\section{Conclusions}\label{sec:5}
We have investigated whether the confining quark potential in quenched QCD is caused by CDG gauge-invariant monopoles. Our main novelty is to construct the CDG colour field $n^j = \theta \lambda^j \theta^\dagger$ from the eigenvectors of the Wilson loop, which allows us to access the full symmetry group of the monopoles and permits a theoretical discussion. Our numerical results suggest that the restricted field strength is dominated by peaks one lattice spacing across and that these peaks are responsible for the confining potential. We see some, though not yet convincing, evidence supporting the theoretical expectation that these peaks are close to the point where the SU(2) subcomponents of $\theta$ are off-diagonal and that the $\theta$ field winds around these peaks. This suggests that these monopoles are at least partially responsible for confinement. We will expand this argument and give full details of the calculation in a subsequent publication.
 \acknowledgments
Computer calculations were carried out on servers at Seoul National University. Funding was provided by the BK21 program of the NRF, Republic of Korea. The research of W.~Lee is supported by the Creative Research
Initiatives Program (2012-0000241) of the NRF grant funded by the
Korean government (MEST).
W.~Lee would like to acknowledge the support from KISTI supercomputing
center through the strategic support program for the supercomputing
application research [No. KSC-2011-G2-06].
 YMC is supported in part by NRF grant (2012-002-134)
funded by MEST.
\bibliographystyle{JHEP_mcite}
\bibliography{weyl}

\end{document}